\documentclass[aps,twocolumn,floatfix,showpacs,amssymb]{revtex4}
\usepackage{graphicx,bm}
\usepackage{hyperref}
\usepackage{amsmath}
\usepackage{mathbbol}
\usepackage{color}

\begin{document}

\title{One-dimensional ultracold atomic gases: Impact of the effective range on integrability}
\author{Tom Kristensen$^{1,2}$ and Ludovic Pricoupenko$^{1,2}$}
\affiliation
{
1- Sorbonne Universit\'{e}s, UPMC Univ Paris 06, UMR 7600, Laboratoire de Physique Th\'{e}orique de la Mati\`{e}re Condens\'{e}e, 
F-75005, Paris, France\\
2- CNRS, UMR 7600, Laboratoire de Physique Th\'{e}orique de la Mati\`{e}re Condens\'{e}e, F-75005, Paris, France
}
\date{\today}

\begin{abstract}

Three identical bosons or fermions are considered in the limit of zero-range interactions and finite effective range. By using a two channel model, we show that these systems are not integrable and that the wave function  verifies specific continuity conditions at the contact of three particles. This last feature permits us to solve a contradiction brought by the contact model which can lead to an opposite result concerning the integrability issue. For fermions, the vicinity of integrability is characterized by large deviations with respect to the predictions of the Bethe ansatz. 
\end{abstract}


\maketitle

\section{Introduction}

The one-dimensional (1D) Bose and Fermi gases with zero-range interactions are celebrated examples of exactly solvable many-body problems \cite{Lie63,McG64,Fuc04,Sut07}. Ultracold atoms offer the stupendous possibility to achieve these systems in the degenerate regime by using highly elongated cigar traps \cite{QGLD,Blo08,Gua13}. Moreover, using magnetic Feshbach resonances and/or tuning the trap parameters make it possible to study 1D systems in strongly correlated regimes. This way, the Tonks-Girardeau and the super Tonks-Girardeau phases have been achieved \cite{Gir60,Ast05,Kin04a,Par04,Hal09}. In addition, the existence of confinement induced resonances and resonances shifts have been confirmed \cite{Gun05,Hal10}.  
Analogously to the three-dimensional (3D) case \cite{Pet04b,Jon08,Pri13}, it has been shown recently that considering a large 1D effective range parameter permits one to enrich the phase diagram of the Bose and Fermi 1D degenerate gases \cite{Ima10,Qi13}. For bosons, this regime is achieved for narrow resonances i.e., in the limit of small Feshbach coupling between atoms and diatomic molecules \cite{Kri15a}. For fermions, the ${p}$-wave scattering resonance is intrinsically narrow and this regime can be easily reached \cite{Pri08}. 
As in 3D systems, for small energy processes, the regime of large effective range can be studied in the limit where the actual radius of the interacting potentials is formally zero whereas the effective range and the scattering length of the model are finite. 
Using a contact model (CM), it was shown for one-component fermions in Ref.~\cite{Ima10} and for identical bosons in Ref.~\cite{Qi13} that the eigenstates of these systems are given by the Bethe ansatz (BA) and are thus integrable in the limit of large effective range \cite{Perturbative}. This result is in strong contradiction with the McGuire-Yang-Baxter criterion which when applied in this regime shows diffractive effects in multiple scattering \cite{McG64,Yur08}. Hence integrability is inherently not possible.

In this paper, we consider the three-body problem, which is intimately related to the integrability issue \cite{Sut07}. To this end we use a Hamiltonian two-channel model (HTCM), which encapsulates the Feshbach mechanism. Whereas the CM and  the HTCM are strictly equivalent at the two-body level, in the three-body problem the HTCM gives large deviations with respect to the predictions based on the BA. We show that in the limit of the contact of three particles, all the solutions of the HTCM have the same type of singularity not satisfied by the BA. The behavior of the wave function in the limit where the three particles fall one on top of the other appears then as a key ingredient in the violation of the integrability. We show that equivalence of the CM and of the HTCM can be achieved at the three-body level by imposing continuity conditions on the wave function.

\section{Two-body problem and McGuire-Yang-Baxter criterion}
 
Our modeling of the system is based on a parameterization of the two-body 1D asymptotic scattering states including the effective range term. For an incoming wave of relative wave number ${k_0}$ and relative coordinate ${z}$, we write it as
\begin{equation}
\langle z | \psi_{k_0} \rangle = e^{ik_0z} + \big[ f_{0}(k_0) + {\rm sgn}(z)  f_{1}(k_0) \big] e^{ik_0|z|} .
\label{eq:scattering}
\end{equation}
In Eq.~\eqref{eq:scattering}, ${{\rm sgn}(z)}$ is the sign function and ${f_{0}}$ (${f_{1}}$) is the scattering amplitude in the even (odd) sector, parameterized as 
\begin{equation}
f_\eta(k_0)=\frac{-(ik_0 a_\eta)^\eta}{1+ik_0 a_\eta + {b}_\eta (ik_0)^{3-\eta}  (-a_\eta)^\eta}.
\label{eq:feta}
\end{equation}
For ultracold atoms in a 1D waveguide, the scattering lengths ${a_\eta}$ and the effective range parameters ${b_\eta}$ in Eq.~\eqref{eq:feta} can be expressed as a function of 3D scattering parameters in the homogeneous space \cite{Ols98,Gra04,Yur05,Kim05,Idz06b,Nai07,Sae12,Pen14}. In what follows, we consider only positive values of the effective range parameter ${{b}_\eta>0}$, an assumption justified in the limit of narrow resonances \cite{Pri08,Kri15a}. From the analyticity of the scattering amplitude, one finds a single bound state, i.e., a dimer ${|\phi_\eta\rangle}$ of energy ${-\hbar^2 \kappa_\eta^2/m}$  in the even sector for all values of ${a_0}$ and  in the odd sector only for positive values of ${a_1}$
\begin{equation}
\langle z |\phi_\eta\rangle =    [1-2 \eta \theta(z)] \times  \sqrt{\frac{\kappa_\eta}{1+2{b}_\eta \kappa^{3-2\eta}_\eta}} \times e^{-{\kappa_\eta} |z|} .
\label{eq:dimer}
\end{equation}
In Eq.~\eqref{Eq:eq:dimer} ${\theta(z)}$ is the Heaviside function and in the odd sector ${(\eta=1)}$, one recognizes the sign function ${{\rm sgn}(z) = 2\theta(z)-1}$. The dimer binding wave number ${\kappa_\eta}$ in Eq.~\eqref{eq:dimer} is the positive root of 
\begin{equation}
1 - a_\eta \kappa_\eta - {b}_\eta (a_\eta)^\eta \kappa_\eta^{3-\eta} = 0 . 
\label{eq:kappa_eta} 
\end{equation}
We now come to the integrability issue for a system of ${N}$ one-component bosons (fermions) where the two-body scattering occurs only in the even (in the odd) sector \cite{other_cause}. Integrability means that the eigenstates are given by the BA and there is thus no diffractive scattering i.e., the wave numbers of the particles are globally conserved after multiple collisions in the system  \cite{Sut07}. 
The expressions of the transmission ${(t_\eta)}$ and reflection ${(r_\eta)}$ coefficients in the scattering of two identical particles are thus particularly relevant. They are defined by an alternative expression of the asymptotic scattering state in Eq.~\eqref{eq:scattering} where the interaction occurs only in one of the sectors ${\eta=0}$ or ${\eta=1}$:
\begin{equation}
\langle z|\psi_{k_0}\rangle = \left\{
\begin{array}{lc} 
e^{ik_0z} + r_\eta(k_0) \times e^{-ik_0z} &\ \text{for}\ z < 0 \\
 t_\eta(k_0) \times e^{ik_0z} &\ \text{for}\ z > 0
 \end{array} \right. 
\end{equation} 
From Eq.~\eqref{eq:scattering}, considering a pair of particles ${(i,j)}$ of wave numbers ${(k_i,k_j)}$, the transmission and reflection coefficients are related to the exchange of momentum between the scattering particles with:
\begin{align}
&{t_\eta^{ij}= 1+ f_\eta\left(\frac{k_i-k_j}{2}\right)}\\ 
&{r_\eta^{ij}=(-1)^\eta f_\eta\left(\frac{k_i-k_j}{2}\right)} .
\end{align}
A necessary condition for integrability is given by the McGuire-Yang-Baxter criterion, which follows from the absence of diffractive scattering in the three-body integrable problem \cite{McG64,Yur08}:
\begin{equation}
r_\eta^{12} r_\eta^{13} t_\eta^{23} + t_\eta^{12} r_\eta^{13} r_\eta^{23} = r_\eta^{23} t_\eta^{13} r_\eta^{12} .
\label{eq:McGuire_Yang_Baxter}
\end{equation}
For ${\eta=0}$ this last equality is verified  if and only if ${b_0=0}$ (Lieb Liniger model) and for ${\eta=1}$ if and only if ${a_1=0}$ or ${|a_1|=\infty}$ i.e., in  the Fermi Tonks-Girardeau (FTG) regime \cite{Wri05,Gir06a}. 
This is in strong contradiction with the results of Refs.~\cite{Ima10,Qi13} where the BA was used as an eigenstate of contact models in regimes where Eq.~\eqref{eq:McGuire_Yang_Baxter} is not satisfied. To understand this discrepancy,  in the rest of this paper we focus on the three-body problem which has the advantage of the simplicity while being a cornerstone of the integrability. 

\section{Contact model}

\subsection{Contact conditions and pseudo-potentials}

We first use a CM which includes the effective range as a straightforward generalization of the Lieb Liniger model and introduced in Refs.~\cite{Pri08,Kri15a}. It is analogous to the one used in the context of narrow Feshbach resonances for atoms in the three dimensional space~\cite{Pet04b,Pri06a,Pri06b}. For convenience we introduce the shorthand notations ${(z)_N}$ for the ${N}$ coordinates of the system and ${(z_{ij})}$ for the relative coordinate of the pair of particles ${(i,j)}$:
\begin{equation}
(z)_N \equiv (z_1,z_2, \dots z_N), \qquad  z_{ij}=z_i-z_j .
\end{equation}
The center of mass of the pair ${(i,j)}$ is denoted ${(Z_{ij})}$, and the relative distance between the pair and the particle ${(k)}$ is denoted ${(Z_{ij,k})}$:
\begin{equation}
Z_{ij}=\frac{z_i+z_j}{2} \quad ; \quad Z_{ij,k} =\frac{z_i+z_j}{2}-z_k
\label{cdm}
\end{equation}
The CM is defined as follows: firstly, for all the configurations where ${\forall i\ne j}$, ${z_i \ne z_j}$,  the wave function ${\langle (z)_N |\Psi \rangle}$ verifies the Schr\"{o}dinger equation without any interaction between particles; secondly, for each pair of interacting particles ${(i,j)}$ the wave function verifies the contact conditions 
\begin{equation}
\lim_{z_{ij} \to 0^+} \left( 1 + a_\eta \partial_{z_{ij}} +  (-a_\eta)^\eta {b}_\eta \partial_{z_{ij}}^{3-\eta}\right) \langle   {(z)_N} | \hat{\Pi}^{ij}_\eta | \Psi \rangle = 0 
\label{eq:contact}
\end{equation}	
where  for ${\eta=0}$ (for ${\eta=1}$) the operator ${\hat{\Pi}^{ij}_\eta}$ symmetrizes (antisymmetrizes) the state ${|\Psi \rangle}$ in the exchange of the particles ${i}$ and ${j}$:
\begin{multline}
\langle {(z)_N} |\hat{\Pi}^{ij}_\eta | \Psi \rangle =\frac{1}{2}\left[\langle {z_1, \dots z_i, \dots z_j, \dots z_N} |\Psi\rangle\right. \\ 
\left. + (-1)^\eta \langle {z_1, \dots z_j, \dots z_i, \dots z_N} | \Psi\rangle \right].
\label{eq:symmetrization}
\end{multline}
In Eq.~\eqref{eq:contact}, the positions ${Z_{ij}}$ and ${z_k}$ (where ${k\ne i,j}$) are kept fixed \cite{caution}. 
One can verify that the exact expressions of the scattering amplitudes in Eq.~\eqref{eq:feta} are deduced from the contact conditions of Eq.~\eqref{eq:contact} by using the wave-function of Eq.~\eqref{eq:scattering}. 

Another equivalent way to implement the contact model is to include directly the contact condition in the Schr\"{o}dinger equation by using the $\Lambda$ potentials ${\hat{V}_\eta^{ij}(\Lambda)}$ for each pair of interacting particles ${(i,j)}$. For a pair of  particles of reduced mass $\mu$, in the even sector of the interaction:
\begin{multline}
\langle (z)_N | V_{0}^{ij}(\Lambda) | \psi \rangle =  -\frac{\hbar^2}{\mu}  f_{0}(i\Lambda)  \delta(z_{ij})
\times \lim_{z_{ij}\to 0^+}  \bigg[ \Lambda \\ 
+ {(1-\Lambda^3 b_0)} \partial_{z_{ij}} + \Lambda b_0 \partial_{z_{ij}}^3  \bigg]
 \langle (z)_N | \hat{\Pi}^{ij}_{0} | \psi \rangle
\label{eq:Vpseudo_even}
\end{multline}
and in the odd sector of the interaction:
\begin{multline}
\langle (z)_N | V_{1}^{ij}(\Lambda) | \psi \rangle = - \frac{\hbar^2}{\mu} \frac{f_1(i\Lambda)}{\Lambda} \delta'(z_{ij})
\times \lim_{z_{ij}\to 0^+} 
\bigg( \Lambda \\
+\Lambda^2 b_1 + \partial_{z_{ij}} - b_1 \partial_{z_{ij}}^2    \bigg)
 \langle (z)_N | \hat{\Pi}^{ij}_{1} | \psi \rangle .
\label{eq:Vpseudo_odd}
\end{multline}
In Eqs.~\eqref{eq:Vpseudo_even} and \eqref{eq:Vpseudo_odd}, ${\Lambda}$ is an arbitrary parameter, i.e., the action of the pseudo-potential on exact eigenstates do not depend on the value of ${\Lambda}$ \cite{Ols01,Pri08,Pri11a}.

\subsection{Puzzling result for the trimers}

In the regime where a dimer exists and if the system is integrable, then the ground state for three identical particles of mass $m$ is a trimer of energy ${-4\hbar^2\kappa^2/m}$ given by the BA \cite{McG64}
\begin{equation}
\langle  z_1,z_2,z_3 |\psi^{\rm p, BA}_{\eta} \rangle = e^{-\kappa \sum_{i<j} |z_{ij}|} \prod_{i<j}\left[1-2\eta\theta(z_{ij})\right] .
\label{eq:BA_bound}
\end{equation}
Following the standard method in Refs~\cite{Lie63,Cas01}, one considers the contact condition in Eq.~\eqref{eq:contact} for each pair ${(i,j)}$ in configurations where the third particle $k$ is distinct from the center of mass ${Z_{ij}}$ (i.e., ${z_k \ne Z_{ij}}$). For instance, in the case where ${z_2<z_1<z_3}$:
\begin{equation}
\langle  z_1,z_2,z_3 |\psi^{\rm p, BA}_{\eta} \rangle =  e^{-\kappa \left(z_{12}-2 Z_{12,3}\right)} .
\label{eq:BA_bound_example}
\end{equation}
Applying the contact condition for the pair $(1,2)$ on Eq.~\eqref{eq:BA_bound_example} gives ${\kappa=\kappa_\eta}$. The same reasoning for the other configurations give the same result. Moreover, the wave function in Eq.~\eqref{eq:BA_bound} is a solution of the free Schr\"{o}dinger equation almost every where excepted at the contact of two or three particles. Thus surprisingly, the BA for the trimer appears as an eigenstate of the CM with the binding wavenumber ${q_\eta^{\rm t, BA}=2\kappa_\eta}$, in deep contradiction with the McGuire-Yang-Baxter criterion.

\section{The consistency obtained from the two-channel model}

\subsection{Two-channel Hamiltonian}

The consistency of the CM is thus puzzling and to go further we now use a HTCM which is a more conventional approach. 
In this model, the scattering process between two particles is only due to the coherent coupling between the pair of particles and a molecular state of mass ${2m}$. For a plane wave of wave number ${k}$, we choose the convention ${\langle z | k \rangle = \exp(ikz)}$ and we denote the creation operator in the open channel ${\hat{a}_{\eta,k}^\dagger}$, where ${\eta=0}$ for bosons and ${\eta=1}$ for fermions. The creation operator for molecules in the closed channel is denoted by ${\hat{b}_{\eta,k}^\dagger}$, where the index ${\eta}$ permits one to distinguish the composite boson (i.e., the molecule) made of two fermions, from the molecule made of two bosons. We consider only pure systems with identical particles and for each system (${\eta=0}$ or ${\eta=1}$), the Hamiltonian is
\begin{multline}
\hat{H}_\eta =\int \frac{dk}{2\pi} \left[ \epsilon_k \hat{a}_{\eta,k}^\dagger \hat{a}_{\eta,k} +\left( \frac{\epsilon_k}{2} + E_{\eta}^{\rm m} \right) \hat{b}_{\eta,k}^\dagger \hat{b}_{\eta,k} \right]\\
+ \biggl[ \frac{\hbar^2 \lambda_\eta}{m}  \int \frac{dKdk}{(2\pi)^2} \langle k|\delta^\eta_\epsilon\rangle \hat{a}_{\eta,\frac{K}{2}+k}^\dagger \hat{a}_{\eta,\frac{K}{2}-k}^\dagger \hat{b}_{\eta,K} + {\rm h.c.} \biggr]
\label{eq:Hamiltonian-2channel}
\end{multline}
In Eq.~\eqref{eq:Hamiltonian-2channel} ${\epsilon_k=\frac{\hbar^2 k^2}{2m}}$ is the single particle kinetic energy, ${\lambda_\eta}$ is the strength of the coherent coupling between the two channels and ${E_{\eta}^{\rm m}}$ is the internal energy of the molecular state. The function ${\langle k|\delta^\eta_\epsilon\rangle}$ in the second line of Eq.~\eqref{eq:Hamiltonian-2channel} is a cut-off for the inter-channel coupling
\begin{equation}
\langle k|\delta^\eta_\epsilon\rangle  = (ik)^\eta e^{-k^2\epsilon^2/4} .
\end{equation}
Physically, the short-range parameter ${\epsilon}$ represents the length scale below which the collisional properties have a 3D character. For atoms moving in the monomode regime of a 1D harmonic waveguide of atomic frequency $\omega_\perp$, it  is typically of the order of the transverse length ${a_\perp=\sqrt{\hbar/(m\omega_\perp)}}$. At this scale the 1D effective model of Eq.~\eqref{eq:Hamiltonian-2channel} is no more relevant. This explains the fundamental interest of considering the zero-range limit (${\epsilon \to 0}$) which permits one to capture the universal 1D properties for energies much smaller than the level spacing in the waveguide i.e., ${\hbar^2/(ma_\perp^2)}$. In the zero-range limit, the scattering lengths and the effective range parameters of the HTCM are given by
\begin{equation}
a_0 = \frac{m E^{\rm m}_{0} }{\hbar^2 |\lambda_{0}|^2}
; 
a_1 \underset{\epsilon\to 0}{=} \frac{1}{ \sqrt{\frac{2}{\pi}} \frac{1}{\epsilon} - \frac{m E^{\rm m}_1 }{\hbar^2 |\lambda_{1}|^2}}  
;
{b}_{\eta}= \frac{1}{|\lambda_{\eta}|^2}
.
\label{eq:mapping}
\end{equation}
The molecular energy in the odd sector ${(E^{\rm m}_1)}$ is a bare parameter which diverges in the zero-range limit in such a way that ${a_1}$ keeps a desired finite value, whereas the parameters ${E^{\rm m}_{0}}$ and ${\lambda_\eta}$ stay finite in this limit. 

\subsection{Integral equation of the three-body problem}

In the HTCM, a three-body state is the coherent superposition of a particle state (denoted by ${| \psi^{\rm p}_\eta \rangle}$) in the open channel and of a mixed channel state (denoted by ${\sqrt{3!}| \psi^{\rm m}_\eta \rangle/\lambda_\eta}$). In the center of mass frame, it  can be written as
\begin{multline} 
|\Psi_\eta \rangle = \int \frac{dk dK}{(2\pi)^2} \frac{\langle k,K | \psi^{\rm p}_\eta \rangle}{\sqrt{3!}} \hat{a}^\dagger_{\eta,\frac{K}{2}+k} \hat{a}^\dagger_{\eta,\frac{K}{2}-k} \hat{a}^\dagger_{\eta,-K}|0\rangle \\
+ \int \frac{dK}{(2\pi)}  \frac{\langle K | \psi^{\rm m}_\eta \rangle\sqrt{3!}}{\lambda_\eta} 
\hat{b}^\dagger_{\eta,K} \hat{a}^\dagger_{\eta,-K} |0 \rangle .
\label{eq:3b-ansatz}
\end{multline}
For a positive energy ${(E>0)}$, ${|\Psi_\eta\rangle}$  is a scattering state and we denote the three-particle incoming state by  ${| \psi^{\rm 0}_\eta \rangle}$. In Eq.~\eqref{eq:3b-ansatz} ${\langle k,K| \psi^{\rm p}_\eta \rangle}$ is symmetric (for ${\eta=0}$) or antisymmetric (for ${\eta=1}$) in the exchange of two particles i.e., in the transformation ${(k \to -k)}$ or in the transformations ${(k \to k_\pm; K \to K_\pm)}$ where 
\begin{equation}
\left\{
\begin{array}{l}
{k_+ = -\frac{3K}{4} + \frac{k}{2}}\\
{K_+ = -\frac{K}{2} - k}
\end{array}
\right.
\qquad  
\left\{
\begin{array}{l}
{k_- = -\frac{3K}{4} - \frac{k}{2}}\\
{K_- = -\frac{K}{2} + k}.
\end{array}
\right.
\label{eq:def_kpm}
\end{equation}
The projection of the Schr\"{o}dinger equation at energy ${E}$ onto the open channel gives
\begin{multline} 
(E-  \frac{3\epsilon_K }{2} - 2\epsilon_k) \langle k , K | \psi^{\rm p}_\eta \rangle = \frac{2\hbar^2}{m} \left[ \langle k|\delta^\eta_\epsilon\rangle \langle K | \psi^{\rm m}_\eta \rangle \right.\\
\left.
+\langle -k_+|\delta^\eta_\epsilon\rangle \langle  K_+ | \psi^{\rm m}_\eta \rangle 
+\langle k_- |\delta^\eta_\epsilon\rangle \langle  K_- | \psi^{\rm m}_\eta \rangle \right] .
\label{eq:Schrodi_open}
\end{multline}
Combining Eq.~\eqref{eq:Schrodi_open} with the projection of the Schr\"{o}dinger equation onto the one atom plus one molecule space, one obtains in the zero-range limit a 1D Skornyakov Ter-Martirosyan equation \cite{Sko57}
\begin{multline}
\frac{(ik^{\rm rel}_{K})^{2\eta-1} \langle K|\psi^{\rm m}_\eta \rangle}{f_\eta(k^{\rm rel}_{K})} +
\int \frac{dk}{2\pi}  \mathcal M_\eta(K,k,E) \langle k|\psi^{\rm m}_\eta \rangle\\
=\int \frac{dk}{2\pi} (-ik)^\eta \langle k,K | \psi^{0}_\eta \rangle .
\label{eq:1D-STM}
\end{multline}
In Eq.~\eqref{eq:1D-STM} we have introduced the kernel
\begin{equation}
\mathcal M_{\eta}(K,k,E) = \frac{4(k+K/2)^\eta(K+k/2)^\eta} {-\frac{m}{\hbar^2}(E+i0^+) +K^2 + kK +k^2} 
\label{eq:kernel}
\end{equation}
and the relative momentum ${k^{\rm rel}_{K}  = \sqrt{\frac{mE}{\hbar^2} - \frac{3}{4} K^2}}$, where for a negative argument of the square root one uses the standard analytic continuation in scattering theory i.e., ${\sqrt{-q^2}=-i|q|}$. For a state of negative energy ${(E<0)}$, there is no incoming three-particle state ${(| \psi^{\rm 0}_\eta \rangle=0)}$ and the prescription ${E \to E + i0^+}$ in Eq.~\eqref{eq:kernel} can be omitted. For ${b_\eta\ne 0}$, one can deduce from Eq.~\eqref{eq:1D-STM} the large momentum behavior (${|K|\to \infty}$) of the mixed channel wave function solution of the problem as a function of its value at the contact of the three particles:
\begin{equation}
\langle K|\psi^{\rm m}_\eta \rangle  \sim
\frac{8}{3(-2)^\eta } \frac{\langle Z=0^+|\psi^{\rm\, m}_\eta\rangle+\langle Z=0^-|\psi^{\rm m}_\eta\rangle}{b_\eta K^{4-2\eta}} .
\label{eq:large_K}
\end{equation}

\subsection{Mapping with the contact model}

In the zero-range limit, the cut-off function in the configuration space ${\langle z |\delta^0_\epsilon\rangle}$ [${\langle z |\delta^1_\epsilon\rangle}$] tends to the ${\delta}$ distribution denoted by ${\delta^0(z)}$ [to its first derivative ${\delta'(z)}$ denoted by ${\delta^1(z)}$]. Hence Eq.~\eqref{eq:Schrodi_open} gives the singular behavior of the particle wave function at the contact of two particles. For a pair ${(i,j)}$ located at the distance ${Z_{ij,k}=Z_{ij}-z_k}$ from the third particle ${k}$, one has
\begin{equation}
\partial_{z_{ij}}^2 \langle (z)_3 | \psi^{\rm p}_\eta \rangle = 2 \delta^{\eta}(z_{ij}) \langle Z_{ij,k}| \psi^{\rm m}_\eta \rangle + \text{ "non $\delta$ terms"}.
\label{eq:singularity}
\end{equation}
To achieve the mapping with the contact model, one first imposes that the three-body wave function coincides with the particle wave function of the two-channel model. The Schr\"{o}dinger equation obtained with the $\Lambda$ potential must also coincide exactly with the zero range limit of Eq.~\eqref{eq:Schrodi_open}. Thus, the action of the $\Lambda$ potential on a three-particle state verifies
\begin{equation}
\langle (z)_3 | \hat{V}_\eta^{ij}(\Lambda) | \psi^{\rm p}_\eta \rangle = \frac{2\hbar^2}{m} \delta^\eta(z_{ij})  \langle Z_{ij,k}| \psi^{\rm m}_\eta \rangle\ \forall \Lambda
\label{eq:action_Vpseudo}
\end{equation}
By using the particular expressions of the ${\Lambda}$ potentials of Eq.~\eqref{eq:Vpseudo_even} and Eq.~\eqref{eq:Vpseudo_odd}, in the limit ${\Lambda \to 0}$ for ${\eta=0}$ and in the limit ${\Lambda \to \infty}$ for $\eta=1$, one also finds:
\begin{equation}
\langle (z)_N | \hat{V}_{\eta}^{ij}(\Lambda) | \psi \rangle = \frac{2\hbar^2}{m} \delta^\eta(z_{ij})
\lim_{z_{ij}\to 0^+ } \partial_{z_{ij}}^{1-\eta} \langle (z)_N |\hat{\Pi}_\eta^{ij} | \psi \rangle,
\label{eq:source}
\end{equation}
a result which does not depend on the value of $\Lambda$ for states which are solutions of the Schr\"{o}dinger equation in the contact model. In the three-body problem and for the contact model, the mixed channel wave function is thus obtained from the three-particle state with the  following relation:
\begin{equation}
\langle Z_{ij,k}| \psi^{\rm m}_\eta \rangle=\lim_{z_{ij}\to 0^+ } \partial_{z_{ij}}^{1-\eta} \langle z_1,z_2,z_3 |\hat{\Pi}_\eta^{ij} | \psi \rangle .
\label{eq:link_molecule_particles}
\end{equation}
From Eq.~\eqref{eq:link_molecule_particles}, in the case of the BA, we denote the mixed channel state ${|\psi^{\rm m, BA}_{\eta} \rangle}$ and using Eqs.~\eqref{eq:link_molecule_particles} and \eqref{eq:BA_bound}, one obtains 
\begin{equation}
\langle Z |\psi^{\rm m, BA}_{\eta} \rangle =-\kappa_\eta^{1-\eta} \left[1-2\eta\theta(Z)\right]^2 e^{-2\kappa_\eta |Z|} ,
\label{eq:BA_configuration}
\end{equation}
a result which can be obtained also directectly from Eq.~\eqref{eq:singularity} \cite{delta}. Applying a Fourier transform on Eq.~\eqref{eq:BA_configuration}, one finds the momentum representation of the mixed-channel state
\begin{equation}
\langle K | \psi^{\rm m, BA}_\eta \rangle = -\frac{4 \kappa_\eta^{2-\eta}}{4 \kappa_\eta^2+K^2} .
\label{eq:BA_momentum}
\end{equation}

\subsection{Trimers}

\begin{figure}[t]
\includegraphics[width=8cm]{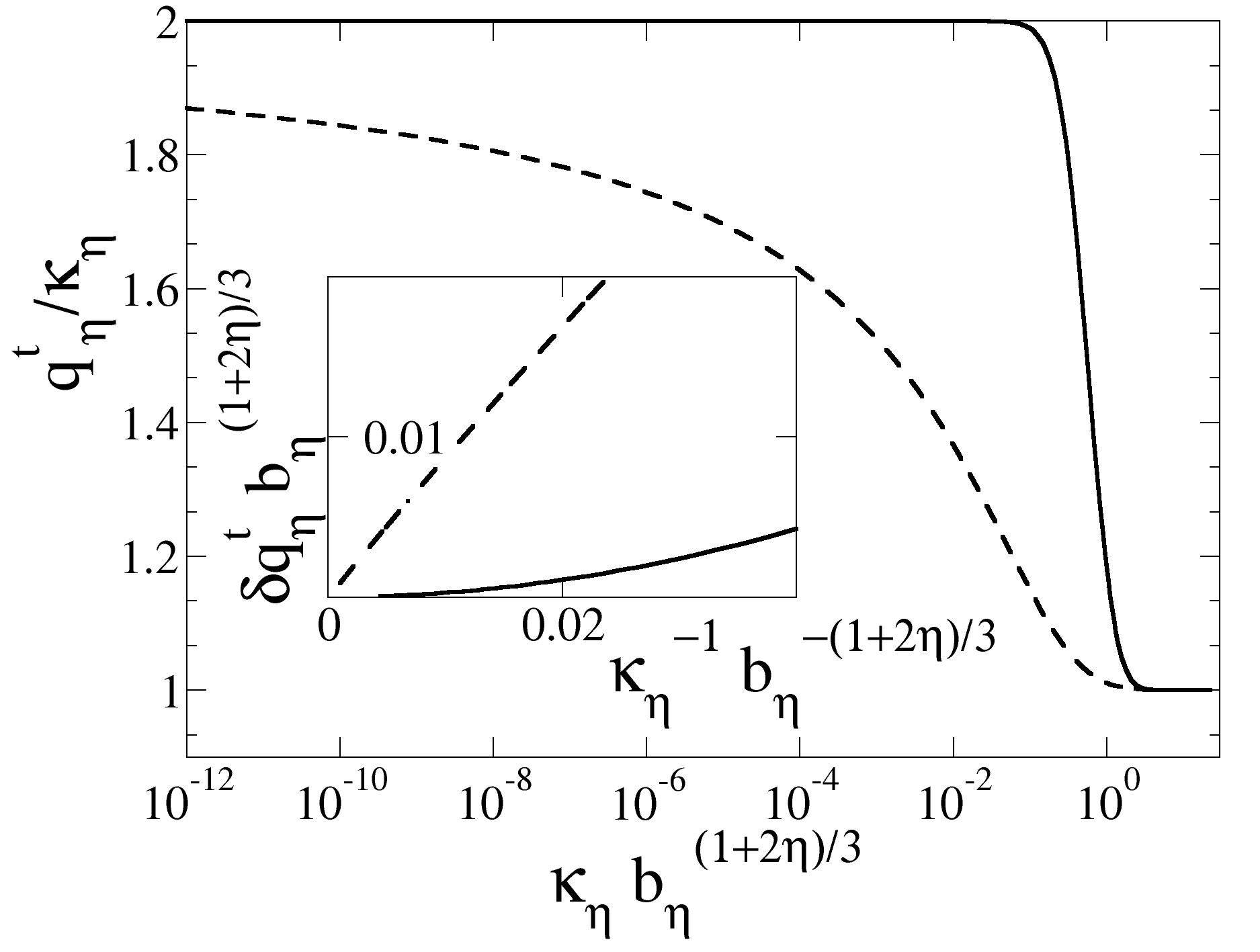}
\caption{Spectrum of the trimers as a function of the wave number of the dimer. Continuous line: bosons (${\eta=0}$), dashed line: fermions (${\eta=1}$). Inset: plot of ${\delta q_\eta^{\rm t}=2 \sqrt{(q_\eta^{\rm t}\,^2 - \kappa_\eta^2)/3}}$ in the region where ${q_\eta^{\rm t} \sim \kappa_\eta}$.}
\label{fig:trimers}
\end{figure}
\begin{figure}[t]
\includegraphics[width=8cm]{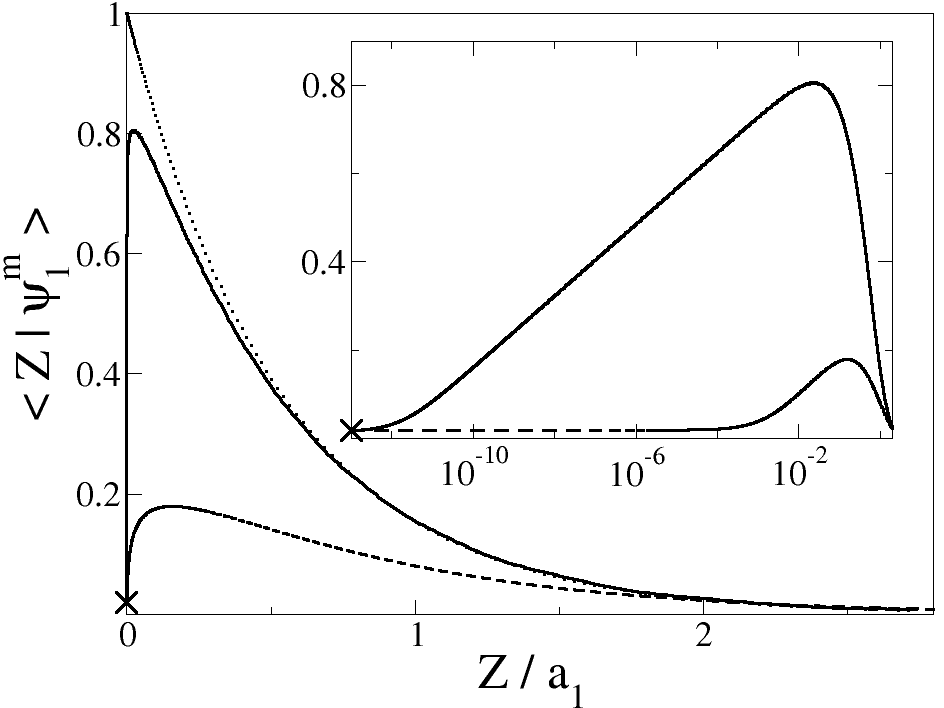}
\caption{Mixed-channel wave function for the fermionic trimer (${\eta=1}$). Solution of Eq.~\eqref{eq:1D-STM} plotted in the configuration space for  ${(a_1=10^{3} b_1)}$ (dashed line)  and ${(a_1=10^{12} b_1)}$ (continuous line). Dotted line: BA mixed-channel state in Eq.~\eqref{eq:BA_configuration} with a normalization chosen for convenience. Inset: same plot where ${Z/a_1}$ is in logarithmic scale. The cross indicates that the solutions of the HTCM do not vanish at the three-body contact.}
\label{fig:trimer_fermion}
\end{figure}

 
We are now ready to compare the trimers obtained from Eq.~\eqref{eq:1D-STM} with the BA. It is clear that for ${b_\eta \ne 0}$, the BA in Eqs.~\eqref{eq:BA_configuration} and \eqref{eq:BA_momentum} does not fulfill the correct asymptotic behavior in Eq.~\eqref{eq:large_K} which confirms the non-integrability. For bosons, this result was found in a model including also the direct particle-particle interaction \cite{Yur06}. The fact that for fermions, the three-body problem is ill-defined when both ${b_1=0}$ and  the numerator of Eq.~\eqref{eq:large_K} is not zero, shows also that the BA in Eq.~\eqref{eq:BA_momentum} can never be an exact solution of Eq.~\eqref{eq:1D-STM} \cite{Trick}. This can be shown as follows: firstly,  for ${\eta=1}$ and ${b_1=0}$ at large momentum, Eq.~\eqref{eq:1D-STM} is scale invariant and the mixed-channel wave function can be searched as a power law: ${\langle K |\psi^{\rm m}_\eta\rangle\propto K^s}$; secondly, the integral in the first line of Eq.~\eqref{eq:1D-STM} is definite at least if ${s<-1}$; thirdly, implementing the limit of large momentum in Eq.~\eqref{eq:1D-STM} one finds ${\langle K |\psi^{\rm m}_\eta\rangle\propto 1/K}$ unless the numerator of Eq.~\eqref{eq:large_K} equals zero, which completes the proof. Similarly to the integrable case, we have found numerically that whenever a dimer exists, there exists also one and only one trimer characterized by an even symmetry (i.e., ${\langle K|\psi^{\rm m}_\eta \rangle = \langle -K|\psi^{\rm m}_\eta \rangle}$). We denote the trimer energy by ${E^{\rm t}_\eta=-(\hbar q^{\rm t}_\eta)^2/m}$. In Fig.~(\ref{fig:trimers}) the wave number ${q^{\rm t}_\eta}$ is plotted as a function of the dimer wave number ${\kappa_\eta}$. In the limit of large scattering length ${(a_\eta\gg b_\eta^{1/(3-2\eta)})}$, the binding wave number of the trimer tends to the integrable limit ${q_\eta^{\rm t} \sim 2 \kappa_\eta \sim 2/a_\eta}$. For bosons, the convergence is fast and one can verify straightforwardly that for ${b_0=0}$, Eq.~\eqref{eq:BA_momentum} is the trimer solution of  Eq.~\eqref{eq:1D-STM}. For fermions, the convergence toward the integrable limit is very slow: one finds the approximate law ${q^{\rm t}_1 \sim 2\kappa_1 [1+1.86/\ln(0.466\times b_1/a_1)]}$. The shape of the mixed-channel wave function converges also slowly toward the BA of Eq.~\eqref{eq:BA_momentum} even for very large values of the ratio ${a_1/b_1}$. In Fig.~\ref{fig:trimer_fermion}, the wave function in configuration space is plotted for the ratio ${a_1/b_1=10^3}$ and ${a_1/b_1=10^{12}}$. Even for this last value, there is a clear distinction between the solution of the HTCM and the BA (dotted line) in the vicinity of the three-body contact. In this last region, the deviation with respect to the integrable solution is large due to the discontinuity of the BA at ${Z=0}$ for ${\eta=1}$ in Eq.~\eqref{eq:BA_configuration}
[${{\rm sgn}(0)=0}$ and thus ${\langle Z=0 |\psi^{\rm m, BA}_{1} \rangle =0}$]. Nevertheless, our numerical solutions of the HTCM show that the discontinuity is asymptotically recovered in the integrable limit. In the opposite limit of a large dimer wave number, ${q^{\rm t}_\eta \sim \kappa_\eta}$ and the mixed-channel wave function tends to the expected results for a shallow two-body (i.e., atom-dimer) bound state ${\langle Z |\psi^{\rm m, BA}_\eta \rangle \sim \exp(-\delta q_\eta^{\rm t}|Z|)}$ and ${\delta q_\eta^{\rm t}= 2\sqrt{(q_\eta^{\rm t}\,^2 - \kappa_\eta^2)/3}}$. We find numerically ${\delta q_0^{\rm t}\sim -2.66/a_0}$ for ${a_0\to -\infty}$ (where ${\kappa_0 \sim \sqrt{-a_0/ b_0}}$)
and ${\delta q_1^{\rm t}\sim  0.835 \times\sqrt{a_1/b_1^3}}$ for ${a_1\to 0^+}$ (where ${\kappa_1 \sim 1/\sqrt{a_1 b_1}}$). 

\subsection{Atom-dimer scattering lengths}

\begin{figure}[t]
\includegraphics[width=8cm]{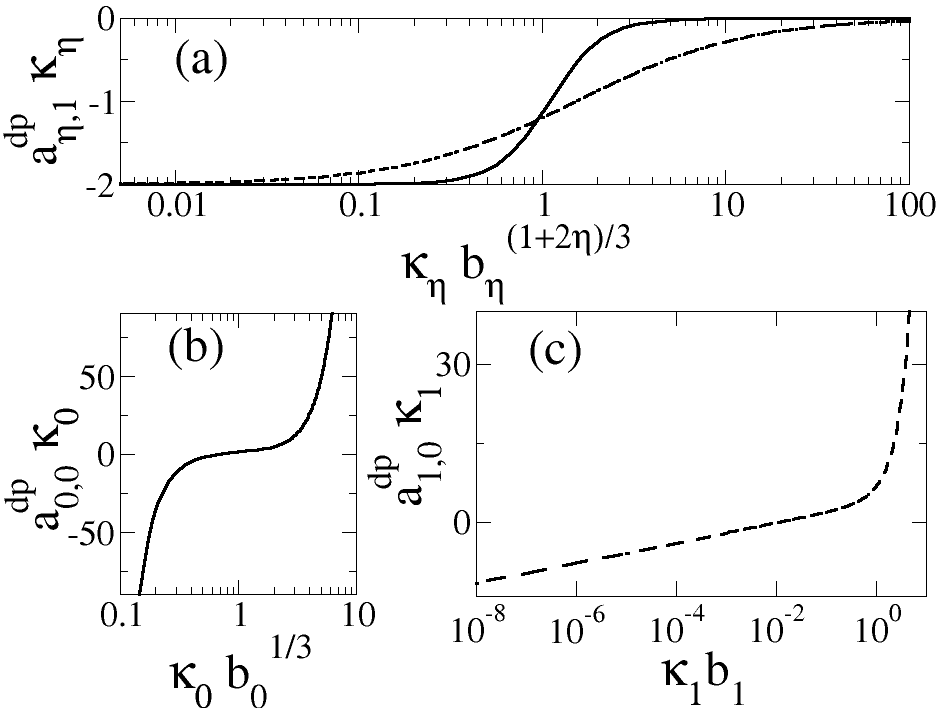}
\caption{Dimer-particle scattering lengths defined in Eq.~\eqref{eq:adp} plotted as a function of the dimer binding wave number. Continuous line: bosonic system ($\eta=0$); dashed line: fermionic system ($\eta=1$). (a): Odd sector ${(\eta'=1)}$; (b): even sector for bosons; (c): even sector for fermions.}
\label{fig:atom-dimer-scattering}
\end{figure}

We have also solved the dimer-particle scattering problem, for an incoming wave of momentum ${k_0}$ and a negative energy:
\begin{equation}
E=\frac{3\hbar^2}{4m} k_0^2 - \frac{\hbar^2 \kappa_\eta^2}{m} <0 .
\end{equation}
The mixed-channel state ${| \psi^{\rm m, scat}_\eta \rangle}$ can be written for an arbitrary overall normalization as
\begin{equation}
\langle K| \psi^{\rm m, scat}_\eta \rangle = 2\pi \delta(K-k_0) + \frac{2ik_0 g_\eta(K,k_0) }{k_0^2+i0^+-K^2} 
\label{eq:ansatz_scatt_atdim}
\end{equation}
and in the configuration space one has for ${Z\ne 0}$
\begin{equation}
\langle Z |  \psi^{\rm m, scat}_\eta \rangle = e^{ik_0 Z} + g_\eta(\text{sgn}(Z) k_0,k_0) e^{ik_0 |Z|} .
\label{eq:scattering_dp}
\end{equation}
The comparison of Eq.~\eqref{eq:scattering_dp} with the generic definition of the 1D scattering amplitudes in Eq.~\eqref{eq:scattering} and Eq.~\eqref{eq:feta} gives the relation between the function ${g_\eta}$ and the even ${(a^{\rm dp}_{\eta,0})}$ or odd ${(a^{\rm dp}_{\eta,1})}$ dimer-particle scattering lengths in the bosonic ${(\eta=0)}$ and fermionic ${(\eta=1)}$ cases:
\begin{equation}
a^{\rm dp}_{\eta,\eta'} =  \lim_{k\to 0} \frac{1-\eta'}{ik} + \frac{g_\eta(k,k) +(1-2\eta')g_\eta(-k,k)}{2ik}
\label{eq:adp}
\end{equation}
Using Eq.~\eqref{eq:1D-STM}, the function $g_\eta(k,k_0)$ is obtained from the integral equation 
\begin{multline}
\frac{(q^{\rm rel}_{K})^{2\eta-1} g_\eta(K,k_0)}{(k_0^2-K^2)f_\eta(iq^{\rm rel}_{K})} -
 {\mathcal P} \int_{-\infty}^\infty\!\! \frac{dk}{2\pi} \, 
\frac{{\mathcal M}_\eta(K,k,E)
}{k_0^2-k^2} g_\eta(k,k_0)\\
- \mathcal M_\eta(K,k_0,E) \frac{g_\eta(k_0,k_0)}{4ik_0}- \mathcal M_\eta(K,-k_0,E) \frac{g_\eta(-k_0,k_0)}{4ik_0}\\=\frac{{\mathcal M}_\eta(K,k_0,E)}{2ik_0},
\label{eq:diff}
\end{multline}
where ${\mathcal P}$ denotes the Cauchy principal value and
\begin{equation}
q^{\rm rel}_{K}=\sqrt{ \kappa_\eta^2 + \frac{3}{4} (K^2-k_0^2) } .
\end{equation}
The four types of particle-dimer scattering lengths are plotted in Fig.~(\ref{fig:atom-dimer-scattering}). In the integrable limit (small values of ${\kappa_\eta}$), the even scattering lengths (${\eta'=0}$) diverge, whereas the odd scattering lengths (${\eta'=1}$) tend toward ${-2/\kappa_\eta}$: a result expected from Ref.~\cite{Mor05a}. In the opposite limit (large values of ${\kappa_\eta}$), the trimer is shallow and the even scattering length is thus given by ${1/\delta q_\eta^{\rm t}}$ [see Fig.~(\ref{fig:trimers})] .
\section{The domain of the contact model}

After this study of the three-body problem with the HTCM, we point out that making the assumption that for ${b_\eta \ne 0}$:
\begin{equation}
{\partial_Z^{1-\eta} \langle Z |\psi^m_\eta\rangle} \ \text{is continuous at} \ Z=0,
\label{eq:continuity}
\end{equation}
filters out unphysical solutions of the CM. For bosons, the continuity of the derivative ${\partial_Z \langle Z |\psi^m_{\eta=0} \rangle}$ at the three-body contact is a necessary condition to avoid the spurious ${K^{-2}}$ behavior at large momentum of ${\langle K |\psi^m_{\eta=0} \rangle}$. For fermions and in the even sector, the condition in Eq.~\eqref{eq:continuity} is necessary  to recover ${\langle Z |\psi^m_{\eta=1} \rangle}$ from  ${\langle K |\psi^m_{\eta=1} \rangle}$ by the inverse Fourier transform for all values of $Z$, including at the three-body contact ${Z=0}$. In the odd sector, a first-order discontinuity at ${Z=0}$ is incompatible with the ${K^{-2}}$ behavior at large momentum of ${\langle K |\psi^m_{\eta=1} \rangle}$ in Eq.~\eqref{eq:large_K}.

The continuity condition in Eq.~\eqref{eq:continuity} excludes the BA from the set of eigenstates of the CM, because it does not belong to the correct domain, whereas it makes it possible to derive Eq.~\eqref{eq:1D-STM} from the CM as follows. Using the standard method, one uses in the Schr\"{o}dinger equation the ${\delta}$ source terms which are related to the two-body singularities of the wave function \cite{Sko57} and correspond to the expressions of the $\Lambda$ potential in Eqs.~\eqref{eq:source} and \eqref{eq:link_molecule_particles}. In our case, one obtains Eq.~\eqref{eq:Schrodi_open} in the exact zero-range limit ${(\epsilon=0)}$. The particle wave function can be then expressed in terms of the mixed-channel wave function with
\begin{multline}
\langle (z)_3 | \psi_\eta \rangle = \langle (z)_3 | \psi_\eta^0 \rangle +  \frac{2\hbar^2}{m}\int \frac{dkdK}{(2\pi)^2} e^{i k z_{12}} e^{i K Z_{12,3} }
\\
\times 
\frac{(i k)^\eta \langle K | \psi^{\rm m}_\eta \rangle 
+(-ik_+)^\eta \langle  K_+ | \psi^{\rm m}_\eta \rangle 
+ (ik_-)^\eta \langle  K_- | \psi^{\rm m}_\eta \rangle }
{E+i0^+-  \frac{3\epsilon_K }{2} - 2\epsilon_k}
\end{multline}
where ${k_\pm,K_\pm}$ are defined in Eq.~\eqref{eq:def_kpm}. The Skornyakov Ter-Martirosyan equation \eqref{eq:1D-STM} follows from the application of the contact condition in Eq.~\eqref{eq:contact} on this last expression.  

\section{Conclusions}

To conclude, we have used the HTCM to study the trimers and the particle-dimer scattering lengths, for three identical particles in one dimension when one takes into account the effective range term. Considering the zero-range limit of the HTCM has been fruitful to define the correct domain of the corresponding 1D CM including the effective range parameter. This way, we conciliate these two different approaches in accordance with the McGuire-Yang-Baxter criterion about the integrability issue. Exploring the phase diagram of the 1D atomic gas from small to large effective ranges is an open issue both experimentally and theoretically. Current experimental techniques make it possible to explore few- and many-body properties in regimes of large effective ranges \cite{Gun05,Hal09,Hal10,Zur12,Zur13}. One expects large deviations from the integrable dynamical properties, observable in the thermalization or in the response functions \cite{Kin06,Rig07,Col11}.


\begin{thebibliography}{99}
\bibitem{Perturbative} In Ref.~\cite{Lee12}, the effect of the effective range in the two-component Bose gas was studied perturbatively using the BA. That study differs thus from the work in Ref.~\cite{Qi13} where the effective range could be arbitrarily large.
\bibitem{other_cause} In our study, the deviation from integrability is analyzed in a strictly 1D context. In Refs.~\cite{Mor05a,Pet12} these deviations are due to the three-dimensional character of the waveguide in quasi 1D ultracold atomic systems.
\bibitem{caution} As in the Lieb Liniger model, one assumes that there is no particle between the particles ${i}$ and ${j}$  when the contact condition for the pair ${(i,j)}$ is performed.
\bibitem{delta} We use the standard identities ${\partial_z|z|={\rm sgn}(z)}$ and ${\partial_z {\rm sgn}(z) = 2\delta(z)}$.
\bibitem{Trick} For ${b_1=0}$, the BA of Eq.~\eqref{eq:BA_momentum} is a solution of Eq.~\eqref{eq:1D-STM} 
if one makes the substitution ${\mathcal M_{1}(K,k,E) \to \mathcal M_{1}(K,k,E)-2}$.
\end{thebibliography}
\end{document}